\documentclass[final,3p,times,twocolumn]{elsarticle}
\usepackage{array}
\usepackage{amssymb}
\usepackage{tikz}
\usepackage{graphicx}
\usepackage{multirow}
\usepackage{tcolorbox}
\usepackage{ragged2e}
\usepackage{float}
\usepackage{lineno}

\begin{document}
\begin{frontmatter}


\author{Amogh Sirnoorkar\corref{cor1}}
\ead{asirnoor@purdue.edu}
\fntext[label2]{Centre for the Advancement of Teaching and Learning of STEM (CATALYST), Purdue University}
\affiliation[inst1]{organization={Centre for the Advancement of Teaching and Learning of STEM (CATALYST), Purdue University},
            addressline={100 N University St},
            city={West Lafayette},
            postcode={47907},
            state={Indiana},
            country={USA}}

\title{Student and AI responses to physics problems examined through the lenses of sensemaking and mechanistic reasoning}

\author[inst2]{Dean Zollman}
\author[inst2]{James T Laverty}
\author[inst3]{Alejandra J Magana}
\author[inst4,inst5]{Sanjay Rebello}
\author[inst1,inst4,inst5]{Lynn A Bryan}


\affiliation[inst2]{organization={Department of Physics, Kansas State University},
            addressline={Cardwell Hall, 1228 N Martin Luther King Junior Drive}, 
            city={Manhattan},
            postcode={66506}, 
            state={Kansas},
            country={USA}}

\affiliation[inst3]{organization={Computer and Information Technology, Purdue University},
            addressline={401 Grant Street}, 
            city={West Lafayette},
             postcode={47907}, 
             state={Indiana},
             country={USA}}

\affiliation[inst4]{organization={Department of Physics and Astronomy, Purdue University},
            addressline={525 Northwestern Avenue}, 
            city={West Lafayette},
             postcode={47907}, 
             state={Indiana},
             country={USA}}

\affiliation[inst5]{organization={Department of Curriculum and Instruction, Purdue University},
            addressline={100 N University St}, 
             city={West Lafayette},
             postcode={47907}, 
             state={Indiana},
             country={USA}}
             
\begin{abstract}
Several reports in education have called for transforming physics learning environments by promoting sensemaking of real-world scenarios in light of curricular ideas. Recent advancements in Generative-Artificial Intelligence has garnered increasing traction in educators' community by virtue of its potential in transforming STEM learning. In this exploratory study, we adopt a mixed-methods approach in comparatively examining student- and AI-generated responses to two different formats of a physics problem through the cognitive lenses of sensemaking and mechanistic reasoning. The student data is derived from think-aloud interviews of introductory students and the AI data comes from ChatGPT's solutions collected using Zero shot approach. The results highlight AI responses to evidence most features of the two processes through well-structured solutions and student responses to effectively leverage representations in their solutions through iterative refinement of arguments. In other words, while AI responses reflect how {\em physics is talked about}, the student responses reflect {\em how physics is practiced}. Implications of these results in light of development and deployment of AI systems in physics pedagogy are discussed.     

\end{abstract}





\end{frontmatter}

\section{Introduction}
\label{sec:intro}

Contemporary reports on higher education have called for shifting the focus of classroom learning from rote-memorization of concepts to sensemaking of real-world scenarios using curricular ideas~\cite{k-12framework}. Consequently, several studies in science education have explored the process of ``making sense of a phenomenon''~\cite{defining,gifford2020categorical} along with developing learning materials that push students to blend everyday and formal ideas~\cite{sirnoorkar2023theoreticaljournal}. While the former group of studies have put forth several theoretical and analytical accounts of sensemaking, the latter have focused on developing research-based assessments to facilitate this process.

Sensemaking is a common cognitive process through which we integrate new information into our knowledge system~\cite{defining}. Through sensemaking, students develop sophisticated epistemology, the feature of making effective transitions between different forms of knowledge (e.g., conceptual, procedural, intuitive, etc.)~\cite{bing2012epistemic}. Sensemaking also leads to generation of new knowledge using existing ideas~\cite{kapon2012reasoning} which in turn, leads into enhanced content understanding~\cite{cannady2019scientific}. Given this significance, there has been an uptick in investigations focused on promoting sensemaking in the learning of physics~\cite{computation,methodssirnoorkar}.    

The Generative-Artificial Intelligence (AI), particularly ChatGPT, has garnered significant traction among educators' community in the last year~\cite{liu2023summary}. The Generative-AI is a Large Language Model that is pre-trained on vast multi-modal data, including texts and images, to generate probabilistic responses.  The chatbot's feature to produce human-like, grammatically sound, and assertive responses to a wide variety of tasks has promised a huge potential in transforming science, technology, engineering, and mathematics (STEM) education. In physics education, studies have focused on exploring the efficacy of AI responses to various forms of assessments that go into a typical course. These include standardized assessments such as Force Concept Inventory~\cite{west2023advances}, free-response homework assignments~\cite{zollman2023analyzing}, clicker questions, multiple choice questions, programming tasks~\cite{kortemeyer2023could}, short-essays assignments~\cite{yeadon2023death}, and end-of-the chapter problems~\cite{wang2023examining}.  

However, a predominant focus of this literature has been on empirically exploring the conceptual accuracy of AI responses to various tasks and discussing the results' implications in light of leveraging AI systems in teaching and learning of physics. However, given the recent efforts to promote authentic knowledge-building practices in physics classrooms~\cite{laverty2022new,sirnoorkar2023theoretical,harris2019designing}, there is also a need to examine the AI responses in light of such practices. The practices-focused examination of AI solutions can shed valuable insights on how AI platforms can complement educators' efforts in facilitating students' scientific inquiry in our classrooms.   

We contribute to the ongoing efforts by going beyond gauging conceptual accuracy of Generative-AI responses to comparatively examining them vis-a-vis students’ responses through the cognitive lenses of sensemaking and mechanistic reasoning. Mechanistic reasoning -- the process of unpacking the underlying mechanism of an observed phenomenon -- too is a central approach to scientific inquiry spanning across multiple disciplines. We adopt a mixed-methods approach in examining students' and ChatGPT's solutions to two versions of a physics problem designed to unpack the underlying mechanism of a real-world context using physics and mathematical principles. While the students' data are derived from think-aloud interviews, the AI data are from ChatGPT's (version 3.5) responses recorded using the Zero Shot approach, i.e., prompting the problems directly to chatbot without providing any context and recording the generated answer. 

In the rest of this manuscript, we answer the following research question: {\em How do students and AI responses compare with each other with regard to the features of sensemaking and mechanistic reasoning?} We address this question through an initial qualitative examination of a sample solution from each data set in terms of sensemaking and mechanistic reasoning followed by the discussion of the quantitative distribution of these markers manifesting across the two sets of data.  

Results highlight the AI responses to reflect the markers of sensemaking and mechanistic reasoning with an exception of the meta-cognitive feature of noticing inconsistencies in ``knowledge system''. However, these features are evidenced despite providing an incorrect reasoning in every response. On the other hand, students data reveal varying degree of evidence of the markers of the two processes but reflecting the rich practices of ``doing physics'', e.g., diagram-based reasoning, iterative refinement of arguments, etc. Thus, we claim that while AI responses reflect {\em how physics is talked about}, student responses reflect {\em how physics is practiced}. Implications of these results, particularly in the context of designing classroom activities involving Generative-AI are discussed.    

This manuscript is structured as follows: in the next section, we briefly review the literature on sensemaking and Large Language Models (LLMs) in physics education. We then present the theoretical constructs of sensemaking and mechanistic reasoning processes in Section~\ref{sec:theory}. In Section~\ref{sec:methods}, we detail the physics problem along with our methods of data collection and analysis. In Section~\ref{sec:qualitative}, we qualitatively analyze a representative solution from each data set before providing the quantitative distribution of the two processes' features in  Section~\ref{sec:quantitative}. We end with discussing the implications and limitations of this work, along with future studies in Sections~\ref{sec:discussion}~and~\ref{sec:conclusion}.

\section{Background}

In this section, we briefly review the literature on sensemaking and the use of Large Language Models (LLMs) in physics education.

\subsection{Sensemaking}

Sensemaking is a common cognitive process through which we integrate new information into our knowledge system. Numerous studies have explored the process of ``making sense of a phenomenon'', particularly in the context of science education. Literature on sensemaking can  be mainly characterized into three groups: theoretical accounts, analytical accounts, and outcomes of the sensemaking process.

The first section of the sensemaking literature involves studies that have explored the theoretical mechanisms involved in the process of ``making sense of a phenomenon''. These studies have described sensemaking in terms of modeling~\cite{sands2020modeling}, transfer~\cite{nokes2013toward}, argumentation~\cite{ford2012dialogic}, an  epistemic frame~\cite{zohrabi2020processes} (in the moment understanding of ``what's going on here''), and an epistemic game~\cite{sensemakinggame} (a systematic approach to inquiry with definite starting and ending points). Given these diverse descriptions of the same process, Odden and Russ provide a coherent account of sensemaking and define it as a process guided by the metacognitive feature of noticing gaps in one's understanding (Refer Section~\ref{sec:theory}). For detailed description on theoretical accounts of sensemaking, we recommend readers the reference~\cite{defining}.    

The second section of the sensemaking literature focuses on the analytical descriptions of the reasoning approaches or snippets of reasoning that qualify as sensemaking. These studies are mainly contextualized in the application of mathematics in physics (also referred to as ``mathematical sensemaking'')~\cite{gifford2020categorical,danielak2014marginalized,kuo2020assessing}. Few analytical descriptions of sensemaking include: seeking coherence between different forms of representations in physics (equations, diagrams, plots, etc.)~\cite{emigh2019equipotential,lenz2020sensemaking}, mapping mathematics with physical world including causal relations~\cite{dreyfus2017mathematical,hahn2019relativity}, and reasoning quantitative problems with qualitative insights~\cite{kuo2020assessing}. For additional details on analytical accounts of sensemaking, refer~\cite{zhao2021development}.

The last group of the sensemaking studies dwell on its outcomes. As noted in Section~\ref{sec:intro}, there are three major outcomes of the sensemaking process: (i) development of sophisticated epistemology, (ii) generation of new knowledge, and (iii) enhanced content understanding. Since sensemaking entails establishing coherence between different forms of knowledge, one of the primary outcomes of the process entails effective navigation between different knowledge forms or sophisticated epistemology~\cite{bing2012epistemic}. Sensemaking also involves leveraging existing ideas through thought experiments, or analogies in making novel claims, thereby leading to the generation of new knowledge- the second outcome of the sensemaking process~\cite{chin2000learning,ruibal2007physics}. Lastly, navigating between different sources knowledge forms in generating new knowledge leads to enhanced content understanding in science~\cite{sirnoorkar2020towards,sirnoorkar2016students,bing2009analyzing} - the final outcome of the sensemaking process.


\subsection{Large Language Models in Physics Education}

The history of technology-enhanced learning in physics dates back to several decades. However, since 2022, there has been a significant increase in the number of investigations focused on Generative-AI in educational contexts mainly driven by the advent of ChatGPT from OpenAI~\footnote{https://snnubiai.github.io/chatgpt\_arxiv\_analysis/}. In physics education, researchers have explored the use of Generative-AI in qualitative education research~\cite{tschisgale2023integrating}, generating synthetic student data~\cite{kieser2023educational}, grading student responses~\cite{kortemeyer2023can}, providing feedback~\cite{wan2023exploring} along with developing and analyzing assessments~\cite{kuchemann2023physics}. However, given the focus of this paper, we primarily focus on the relevant literature involving the intersection of AI and physics assessments.     

The early investigations on Generative-AI in physics education have focused on the AI's efficacy in responding to physics standardized and course assessments. Researchers have explored the increased accuracy of evolving versions of ChatGPT to the Force Concept Inventory. West~\cite{west2023advances} observed the ChatGPT-4 to ace the standardized assessment with an impressive accuracy around 90\%. A similar performance from AI has also been recorded on short-form physics essays where the ChatGPT's responses qualified for a ``First Class'' qualification with an average score of 71$\pm$2\%~\cite{yeadon2023death}.

Several other studies have examined the accuracy of ChatGPT in responding to assessments in typical physics courses and end-of-the chapter problems. Kortemeyer~\cite{kortemeyer2023could} observed ChatGPT to ``just pass'' a typical physics course with various assessment formats including the FCI (60\% accuracy), open-ended homework assignments (55\%), clicker questions (93\%), programming exercises (90\%), and end of the course examinations (47\%). In another study, Yeadon and Halliday~\cite{yeadon2023exploring} analyzed the responses from GPT-4 and GPT-3.5 to 42 examination papers from 10 physics courses between 2018 to 2022 at Durham University. The authors observed the two GPT versions each scoring 50.8\% and 41.6\% respectively on the examination papers with the accuracy dropping on higher-level physics questions requiring complex calculations.

Wang {\em et al}~\cite{wang2023examining} analyze ChatGPT's accuracy in responding to science and engineering problems across well-specified problems (involving abstract contexts and adequate data) and under-specified ones (involving real-world contexts and incomplete data). The authors observe ChatGPT to accurately solve 62.5\% of well-specified problems but only 8.3\% of under-specified problems. Given the varying performance of LLMs on various forms of tasks, Polverini and Gregorcic~\cite{polverini2023understanding} discuss how prompting techniques (ways of posing prompts to the chatbot) can enhance the performance of AI interfaces.  

This broad overview of the literature highlight that the studies on AI in the context of physics assessments have mainly focused on gauging the accuracy of AI responses with little or no focus on the processes that facilitate generation of new knowledge in science. We address this gap in the literature by analyzing students and AI responses to two versions of a physics problem through the cognitive lenses of sensemaking and mechanistic reasoning. 

\section{Theoretical Frameworks}
\label{sec:theory}

In the rest of this section, we discuss the theoretical frameworks of the sensemaking and mechanistic reasoning processes. It is important to note that these two processes are independent of the correctness of solutions. That is, a person may engage in these two processes despite providing an incorrect description or solution of a given scenario.

\subsection{Sensemaking}
\label{subsec:sensemaking}

Sensemaking is a common cognitive process that humans engage in both within and beyond formal settings. It is the process through which we integrate new information into our knowledge system. In the rest of this manuscript, we adopt the following account from Odden and Russ~\cite{defining} who define sensemaking as: 

\begin{quote}
{\em a dynamic process of building or revising an explanation in order to `figure something out' - to ascertain the mechanism underlying a phenomenon in order to resolve a gap or inconsistency in one's understanding. One builds this explanation out of a mix of everyday knowledge and formal knowledge by iteratively proposing and connecting up different ideas on the subject. One also simultaneously checks that those connections and ideas are coherent, both with one another and with other ideas in one's knowledge system.}
\end{quote}

Based on the above definition, we deduce the following markers (henceforth referred as ``sensemaking elements'') which together evidence the sensemaking process:

\begin{enumerate}
    \item Noticing inconsistency in one's understanding.

    \item Use of everyday and formal knowledge while reasoning about a phenomenon. 
    
    \item Ascertaining the underlying mechanism of the phenomenon.
    
    \item Generating and connecting up different ideas.
    
    \item Seeking coherence between the generated ideas.
\end{enumerate}

In terms of the above-mentioned sensemaking elements, we qualitatively analyze the features of sensemaking in student and AI responses to two versions of a physics problem in Section~\ref{sec:qualitative}. We then discuss the quantitative distribution of these elements across the two sets of data in Section~\ref{sec:quantitative}. 

\subsection{Mechanistic reasoning}
\label{subsec:mechanistic reasoning}

Unpacking the underlying mechanism of a phenomenon (the fifth sensemaking element in our list) is a complex cognitive process in itself and has been referred to as ``mechanistic reasoning''. Mechanistic reasoning is a form of causal reasoning that entails the description of the underlying events and factors that give rise to an observed phenomenon~\cite{krist2019identifying,machamer2000thinking}. This form of reasoning often entails generating explanations by transitioning from observable features of a phenomenon at the macro level to the underlying  entities/processes (often at the micro level). The process of ascertaining the mechanism can further involve transitioning back from the micro to the macro features, and testing the validity of the generated explanation by varying the spatial or temporal organization of the entities/processes~\cite{krist2019identifying}.

As an example, consider the phenomenon of a coin placed in a glass of water as being raised when viewed from above. A mechanistic account for this observation can be provided using the principle of refraction of light. When a ray of light passes from a denser medium (water) to a rarer medium (air), it deviates away from an imaginary line perpendicular to the surface between these two media.  This deviation results in the formation of an apparent image of the coin, making it appear closer to the water's surface. Similarly, one can argue that an object in air when viewed from water (e.g., while swimming underwater) would appear to be farther away than the actual distance.

\subsubsection{Analytic framework on mechanistic reasoning}
\label{subsubsec:analytic-framework}

Russ et al.~\cite{russ2008recognizing} put forward an analytic framework to capture the extent of sophistication in mechanistic accounts such as the one described above. This framework characterizes mechanistic reasoning into seven hierarchical codes in the increasing order of their sophistication that relatively evidence mechanistic reasoning. 

\begin{enumerate}
    \item {\em Describing the target phenomenon.} Articulation of a stable phenomenon concerning which reasoning about the underlying mechanism occurs. In our example, the explicit articulation of the apparent shift in the coin's position in water evidences this code.
    
    \item {\em Identifying set up conditions.} Noting the enabling conditions of the environment that trigger regular changes of mechanism leading to the phenomenon. In our case, identifying water as a denser medium, and air as a rarer medium evidences this code.

    \item {\em Identifying entities.} The identification of objects or entities such as a ray of light and the two media that play crucial role in the manifestation of the phenomenon.

    \item {\em Identifying activities.} The identification of actions and interactions of the entities (such as the light ray moving from one media to another) in a mechanism correspond to this code.

    \item {\em Identifying properties of entities.} The identification of general properties of entities (e.g., light traveling in straight lines) that are instrumental in the mechanism. 
    
    \item {\em Identifying organization of entities.} The description of spatial organization, location, and the structure of entities, such as the location of coin in denser medium, reflects this code.

    \item {\em Chaining: backward and forward.} Lastly, reasoning about the events that were crucial for the current behavior of systems (backward chaining), or about the potential future behavior of systems given its current state (forward chaining) mark this code. In our case, an object in air being viewed from water appearing to be farther than the actual distance.

\end{enumerate}

The authors further note that the presence of higher codes (e.g., identifying properties of entities) often subsume the evidence of lower codes (e.g., identifying entities) in mechanistic accounts. Furthermore, contemporary literature has increasingly highlighted the role of representations such as diagrams and gestures during mechanistic reasoning~\cite{de2022students}. Thus, we examine our student data by taking into account the above-mentioned seven codes along with diagrams and gestures employed in the mechanistic explanations. 

\section{Methods}
\label{sec:methods}

\subsection{Context}
The objective of this study is to examine sensemaking and mechanistic reasoning in students and AI responses to a physics problem -- the ``Gravitron problem'' -- posed in two different formats. The problem consists of a rotating cylindrical amusement ride in which riders lean against the wall and students are asked to determine whether riders would fall off the Gravitron's walls under the specified parameters. One of the ways to approach this problem is by first identifying the forces acting on the rider and noting that riders would stay put against the wall if the force of friction would balance the force of gravity. If this condition mathematically holds true, then riders would sustain inside the Gravitron or else they would slide off. A detailed solution of this problem can be found in our earlier work~\cite{sirnoorkar2023sensemaking}. 

\begin{figure}[tb]
    \centering
    \begin{tcolorbox}
     \justify{You are asked to design a Gravitron for the county fair, an amusement park ride where the  rider enters a hollow cylinder, radius of $4.6$~m, the rider leans against the wall and the room spins until it reaches angular velocity, at which point the floor lowers. The coefficient of static friction is $0.2$. You need this ride to sustain mass between $25$-$160$~kg to be able to ride safely and not slide off the wall. If the minimum $\omega$ is $3$~rad/s, will anyone slide down and off the wall at these masses? Explain your reasoning using diagrams, equations and words.}   
     \end{tcolorbox}
    \caption{Open-ended version of the Gravitron task}
    \label{fig:open-ended-Gravitron}
\end{figure}

This problem had two different formats: open-ended (Fig~\ref{fig:open-ended-Gravitron}) and scaffolded (Fig~\ref{fig:scaffolded-Gravitron}). Unlike the open-ended version, the scaffolded version explicitly asked for the underlying assumptions and the free body diagram (diagram highlighting the forces along with their directions acting on the riders) required to solve the problem.  

\subsection{Data}
Our data consists of two sets: student and AI responses. Student responses were collected through two sets of think-aloud interviews each involving 10 introductory students solving a set of physics problems. While the first set of interviews were conducted in the spring of 2018, the second were conducted in Spring 2019. Both interview protocols involved asking students to think-aloud while solving the problems and treat the problem-solving exercise as an untimed exam. The interviewer intervened during instances of prolonged silence through phrases such as “What are you thinking?”  to nudge participants to think-aloud. Participants were provided with a calculator and an equation sheet, and were compensated with a \$20 gift card. Participants' demographic information was not collected during the interviews.   

The problem sets across the two interviews were designed using the “Three-Dimensional Learning Assessment Protocol'' (3D-LAP)~\cite{3DLAP} to elicit scientific practices. While a subset of problems in the first interview were designed to elicit the practice of ``Developing and Using Models'', the second set consisted of problems designed to elicit the practice of ``Using Mathematics”. Two interviews across the each set had issues with audio/video. Thus, the data reported in this study consists of responses from a total of 16 student interviews evenly split across two sets.

On the other hand, the AI data involves responses collected from ChatGPT-3.5 using Zero Shot approach. This method of prompting involves seeking a response from the interface posing the problem directly without providing any prior context. To maintain parity across the student data, ChatGPT was prompted with the physics problem sixteen times (eight each across open-ended and scaffolded versions) by initiating as many separate ``conversations''.

\begin{figure}[tb]
\centering
\begin{tcolorbox}
\justify{You are asked to design a Gravitron for the county fair, an amusement park ride where the enters a hollow cylinder, radius of 4.6 m, the rider leans against the wall and the room spins until it reaches a specified angular velocity $\omega$, at which point the floor lowers. The coefficient of static friction is 0.2. You need this ride to sustain mass between 25-160 kg (i.e., they should be able to ride safely and not slide off the wall. 

A.) What assumptions do you need to make to be able to solve this? 

B.) Create a free body diagram for the rider when the room is spinning. Note all applicable forces and label them. 

C.) If the floor drops out when $\omega$ is 3 rad/s, will anyone slide off the wall in the given mass range? Explain your reasoning.}   
\end{tcolorbox}
\caption{Scaffolded version of the Gravitron task}
\label{fig:scaffolded-Gravitron}
\end{figure}


\subsection{Data Analysis}
As part of the analysis, students' verbal arguments and written solutions to the Gravitron task were analyzed  through the lens of the five sensemaking elements described in Section~\ref{subsec:sensemaking}. The first element -- noticing inconsistencies -- was identified in the responses by noting puzzling questions~\cite{odden2019vexing}, intermittent pauses~\cite{sirnoorkar2023sensemaking}, incomplete arguments, etc. The second sensemaking element -- blending everyday and formal knowledge -- was identified when the responses reflected anamalgamation of formal physics principles with the Gravitron's ``physical system''. Arguments such as ``the rider being pulled down by the gravity'' or ``friction preventing the rider from slipping down'' were coded as the blending of everyday and formal knowledge. The third and fourth elements were captured based on the generation and validation of intuitive, conceptual, or procedural ideas. 

We captured the fifth sensemaking element (mechanistic reasoning) by identifying the nine hierarchical codes described in Section~\ref{subsubsec:analytic-framework}. The first code, description of the target phenomenon, was marked when the responses explicitly noted the phenomenon of riders being held against the wall without slipping. The ``identifying set up conditions'' code was marked when responses highlighted the features such as the lowering of the Gravitron's floor or the spinning of the room. Codes associated with entities (identifying entities, their activities, properties, and organization) were noted by the description of the interplay of the forces on the rider e.g., gravity acting downwards, mass remaining constant, etc. Lastly, the chaining code was marked when the required conclusion was made.

We highlight that our data across student and AI responses have two different modalities. While the student data entails verbal arguments, and written solutions, the ChatGPT's data corresponds to written texts. We acknowledge this major limitation in our methods of data collection/analysis in comparing two different categories of data.  Yet, we adopt this approach as our primary motivation is to empirically explore how students typically reason about a real-world scenario using physics principles and how AI would respond to the same context. As practitioners, we were aware that students would not prefer writing detailed reflective arguments in their solutions, particularly when answering a series of complex problems like the described in this paper.  

\section{Qualitative analysis of a student's and ChatGPT's solution}
\label{sec:qualitative}

In this section, we provide a detailed qualitative analysis of a student named Joshua's (pseudonym) and an AI's solution to the scaffolded version of the Gravitron task. While Joshua's attempt involves an unsuccessful bid to solve the problem, the AI solution provides a detailed but incorrect solution to the problem. The student's solution approach reflects all features of the sensemaking process and most features of mechanistic reasoning (except the ``chaining'' code). The AI's solution  on the other hand, evidences most of the sensemaking features (with an exception of noticing gaps in its ``knowledge system'') and all analytical codes of mechanistic reasoning. Tables~\ref{tab:sensemaking-qualitative}~and~\ref{tab:mechanistic-reasoning-qualitative} highlight the evidentiary quotes from the two data samples highlighting the features of sensemaking and mechanistic reasoning.

We focus on the scaffolded version, since a detailed case-study analyses of two students' sensemaking about the open-ended version of the same task can be found in our earlier work~\cite{sirnoorkar2023sensemaking}. 
As noted in Section~\ref{sec:methods}, the scaffolded version (Figure~\ref{fig:scaffolded-Gravitron}) has three sub-prompts. The first two sub-prompts ``A'' and ``B'' respectively ask about the assumptions required to solve the problem and a free body diagram highlighting the forces acting on the rider. The third sub-prompt ``C'' asks whether riders would fall off the Gravitron's walls under the specified parameters. 


\subsection{Joshua's attempt at the Scaffolded Gravitron task}

\begin{figure}
    \centering
    \includegraphics[scale=0.3]{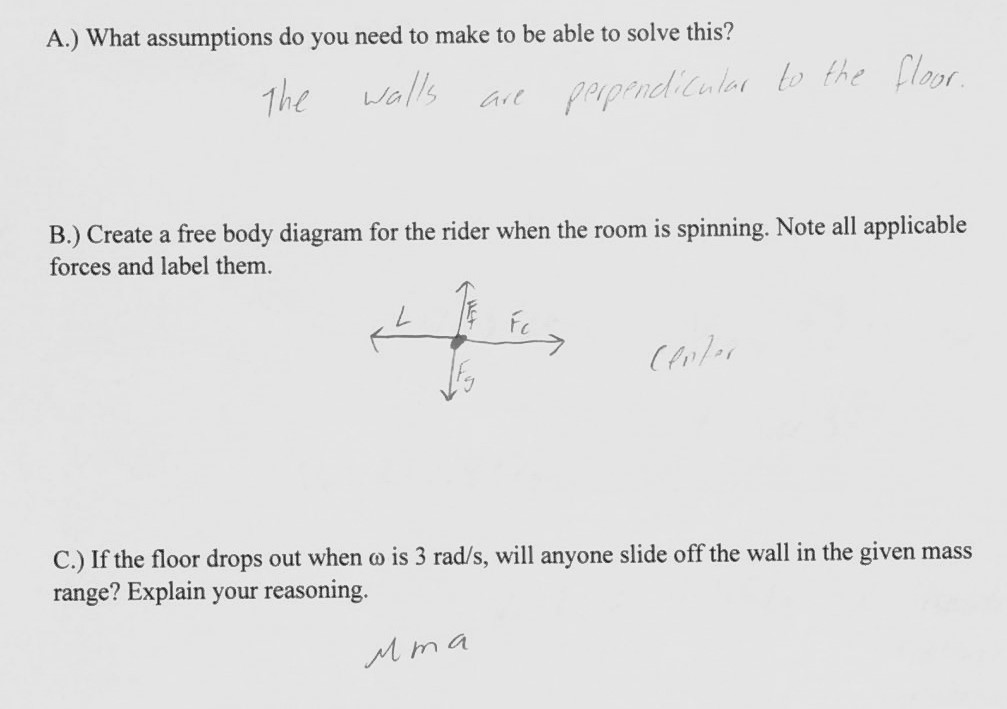}
    \caption{Joshua's written solution to the scaffolded version of the Gravitron task.}
    \label{fig:joshua-solution}
\end{figure}

In response to the first two prompts of the task, Joshua assumes the Gravitron's walls to be perpendicular to its floor and constructs a free-body diagram highlighting the forces acting on the rider. While answering to the first two prompts, he notices a gap in his understanding about a force that counteracts the centripetal force. He tries to address this perceived gap through several conceptual ideas such as angular momentum, impulse, and velocity. Unable to make a breakthrough, he finally quits the problem. Figure~\ref{fig:joshua-solution} highlights Joshua's written solution. Overall, Joshua's attempt reflected all features of the sensemaking process and six of the seven codes of mechanistic reasoning (with the exception of chaining). Tables~\ref{tab:sensemaking-qualitative}~and~\ref{tab:mechanistic-reasoning-qualitative} highlight the representative quotes from Joshua evidencing the sensemaking elements and codes of mechanistic reasoning. These markers are evidenced despite the participant not solving the problem completely. 

Below, we discuss Joshua's attempt into three episodes with the first episode focusing on the problem's sub-parts A and B (focused on assumptions and free-body diagram) and the remaining two episodes focused on the problem's sub-part C. Across each episode, we first summarize the specific segment, present the participant's transcript, and then explicitly call out the features of the sensemaking and mechanistic processes.

\subsubsection{Episode 1. Joshua's assumptions and free-body diagram}

\begin{figure}
    \centering
    \includegraphics[scale=0.15]{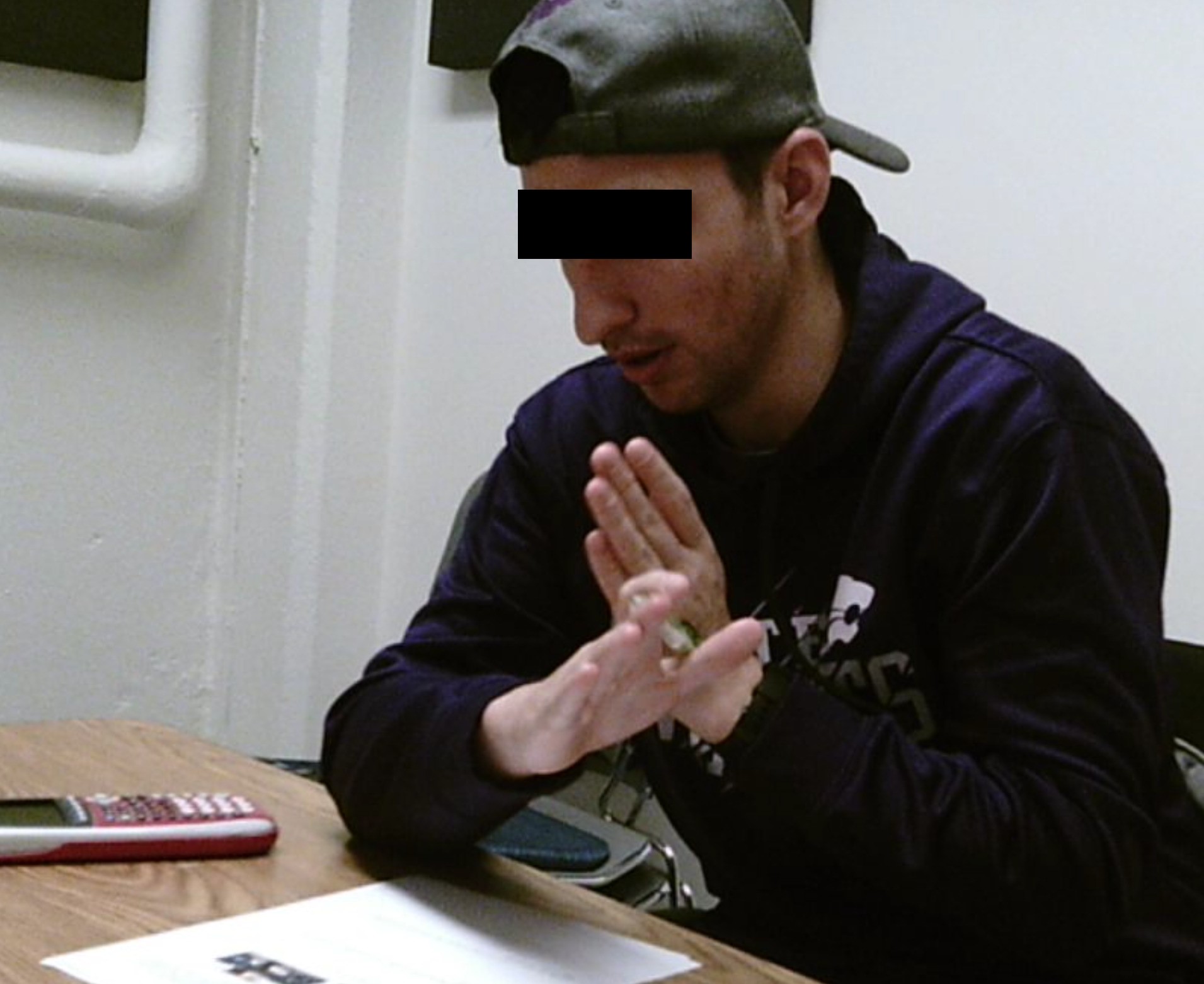}
    \caption{One of the gestures made by Joshua while solving the scaffolded version of the Gravitron task.}
    \label{fig:joshua-gesture}
\end{figure}

{\bf Summary} Upon going through the problem statement, Joshua  articulates that one of the assumptions required to solve the problem should be considering the Gravitron's floor to be orthogonal to its floor. The articulation of this assumption is accompanied by a hand-gesture as shown in Figure~\ref{fig:joshua-gesture}. Moving on to the part B, he constructs a free-body diagram highlighting the rider's center of mass, gravitational force, centripetal force, and angular momentum as shown in Figure~\ref{fig:joshua-solution}.

{\bf Transcript}

\begin{quote}
[Reads prompt A] {\em The angle of the wall... like.. when spinning should be orthogonal? [gestures hand] I mean to the wall with respect to the floor } [inaudible].  {\em I don't know. I am not given theta and I don't know if I need to solve for it. It's going to be pretty difficult for me to do.} [Looks at the equation sheet.]  {\em Well, will come back at it later.} [Writes the assumption].  

[Reads prompt B] {\em Okay, I have the center of gravity} [draws a dot], {\em and centripetal force? And... uh...} [stares at the solution for 15 seconds] {\em I don't know how to solve this. This is a tough question. I don't even know what the centripetal force formula is} [looks at the equation sheet]. {\em I have not looked at any of these recently either. It has to be angular momentum right?} [Draws the left arrow and denotes it as ``L"] {\em I don't exactly remember about the angular momentum.} 
\end{quote}

{\bf Features of sensemaking.} Joshua's approach in the first two sub-prompts highlight his noticing of an inconsistency about his understanding about the forces acting on the Gravitron's rider. His consistent acknowledgement about his (lack of)  understanding about the forces is evidenced by the ``I don't know'' phrases occurring several times while constructing the force diagram. Such acknowledgement of gaps in understanding mark the initiation of the sensemaking process~\cite{odden2019vexing,sensemakinggame}.  

{\bf Codes of mechanistic reasoning.} On the mechanistic reasoning front, this segment reflects the {\em identifying entities and identifying organization of entities} in the phenomenon of the riders being held up inside the Gravitron. Joshua identifies several entities such as Gravitron's walls, its floor, gravitational and centripetal forces, and angular momentum. In addition to noting them, he reasons about their spatial organization through the construction of the free-body diagram and through the verbal arguments such as ``{\em The angle of the wall .. should be orthogonal?}''.

\subsubsection{Episode 2: Joshua's approach to prompt C}

{\bf Summary} Moving to the final sub-part of the problem (Part C), Joshua continues to reason about the forces acting on the rider. Identifying friction as the force counteracting gravity, he struggles to reason about the force counteracting the centripetal force and directed into the wall.   

{\bf Transcript}
\begin{quote}
[Reads the third prompt] {\em So, we need the momentum to outweigh.... What's pointing up? Is it the friction? What's the friction?} [Draws the upper arrow and denoted it as ``$F_f$'']. {\em Okay. So. I haven't entirely figured out the normal force into the wall. I have to come back to it later.} [Looks at the equation sheet] {\em Okay, so the force of friction is keeping them off the ground from Fg. Yeah I don't know.} [inaudible]. {\em Okay, we need this centripetal force as 0.2 times the mass and the acceleration into the wall} [inaudible] {\em which is not the momentum? I do not have anything pointing at the wall right now. I need to figure out how to do that.} 

[Stares at the equation sheet for 25 seconds] {\em Now I am trying to remember how exactly to... I am seriously trying to understand centripetal force right now. Because I only have something pointing to the center of the circle but I don't know what's going out of the circle.} {\em That is a moving mass} [inaudible].  {\em Normal force is equal to m w r which is going to keep them up. Which is equal to mass times acceleration into the wall, which I don't know. Okay} [Stares at the solution for 30 seconds] 
\end{quote}

{\bf Features of sensemaking.} This segment of Joshua's transcript reflects his continued acknowledgement about the gap in his understanding about the forces and blending everyday and formal knowledge to address the perceived gap. While he identifies friction as opposing the gravity (``{\em Okay, so the force of friction is keeping them off the ground from Fg}''), he struggles to reason about the force which is directed into the wall, acting against the centripetal force. This is evidenced by his arguments ``{\em I do not have anything pointing at the wall right now. I need to figure out how to do that.}'' and ``{\em Because I only have something pointing to the center of the circle but I don't know what's going out of the circle.}'' These arguments also reflect the amalgamation of the description of physical features of the system using formal conceptual ideas, the second sensemaking element.

{\bf Codes of mechanistic reasoning.} The same segment of the transcript also evidences several mechanistic reasoning codes including {\em identifying entities, their properties, their activities} and {\em their (spatial) organization.} Joshua identifies frictional and normal forces as entities influential in describing how the rider sustains inside the Gravitron. He further goes on to identify the properties of those forces (entities) such as the normal force being equal to the product of $m, w,$ and $r$ along with the activities engaged by the entities (friction keeping the riders from falling and the constant motion of the rider's mass). Lastly, his arguments also evidence the spatial organization of the entities such as the centripetal force being directed towards to the center of the Gravitron, frictional force directed opposite to gravity, etc.     

\subsubsection{Episode 3: Joshua quits the prompt C }

{\bf Summary} A prolonged pause at the end of the previous segment makes the interviewer intervene with  ``What are you thinking'' to which Joshua replies as trying to figure out the force directed into the wall. He invokes the concepts of impulse, velocity, and momentum in trying to identify the inward-acting force. Unable to make inroads, he quits the problem solving exercise.    

\begin{quote}
[Interviewer] {\em What are you thinking?} [Joshua] {\em I don't know what I am doing. I need to somehow say... uh. I need to figure out how much of their force is heading into the wall using their momentum} [gesturing hand] {\em momentum of the lightest person, but I don't know how to do that. Uh. I don't remember how to do momentum problems. Um, trying to remember about impulse, how we calculated velocity. Force is change in momentum over change in time. But I don't know how much time. And.... I am just thinking whether there is anything wrong in it. I need to find the acceleration into the wall. Wasted a lot of time probably on this. Will move to the next problem}    
\end{quote}

{\bf Features of sensemaking.} The final segment of Joshua's transcript highlights the generation of various ideas and attempts to seek coherence between them while trying to identify the force directed into the Gravitron's walls. He invokes the construct of impulse, trying to remember the calculation of velocity, and also refers to the mathematical definition of force as the rate of change of momentum. Unable to seek a coherence between these ideas, he finally abandons his pursuit.  

{\bf Codes of mechanistic reasoning.} Joshua's final arguments reflect the codes of {\em identifying additional entities} such as impulse, velocity, and time.

\subsection{AI's solution to the Scaffolded Gravitron task}

\begin{figure}
    \centering
    \includegraphics[scale=0.9]{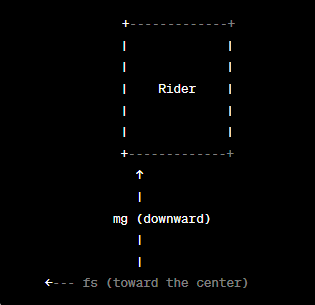}
    \caption{ChatGPT's free-body diagram generated in response to the scaffolded version of the Gravitron task.}
    \label{fig:ai-fbd}
\end{figure}

In the rest of this subsection, we analyze one of the eight AI's solutions to the scaffolded version of the Gravitron task through the lens of sensemaking and mechanistic reasoning. Similar to categorization of Joshua's transcript, we divide the solution into three segments: the first one focusing on the sub-parts A and B,  and the last two segments focusing on the sub-part C. For each segment, we begin with a brief summary, followed by the transcript, and its discussion in terms of the features of sensemaking and mechanistic reasoning.

The AI's transcript explicitly evidences most features of sensemaking with an exception of noticing gaps in its ``knowledge system'' and all the codes of the mechanistic reasoning. Tables~\ref{tab:sensemaking-qualitative}~and~\ref{tab:mechanistic-reasoning-qualitative} highlight the representative quotes from the AI's solution  evidencing sensemaking and  mechanistic reasoning.

\begin{table*}
\begin{center}
\label{tab:sensemaking-qualitative}
\caption{Representative evidentiary quotes of sensemaking elements in Joshua's and ChatGPT's responses to the scaffolded Gravitron task.}
\begin{tabular}{p{0.2\linewidth}p{0.36\linewidth}p{0.36\linewidth}}
\hline
Sensemaking elements & Joshua & ChatGPT \\
\hline 

Noticing inconsistencies in understanding & ``{\em I only have something pointing to the center of the circle but I don't know what's going out of the circle.}''  [Episode 2]  & -  \\

Use of everyday and formal knowledge & ``{\em Normal force is equal to m w r which is going to keep them up.}'' [Episode 2]  & ``{\em If fs\_max is greater than or equal to mg, the riders will stay against the wall and not slide off.}'' [Segment 2]  \\

Generating and connecting ideas & ``{\em Um, trying to remember about impulse, how we calculated velocity. Force is change in momentum over change in time.}'' [Episode 3]  & ``[..]{\em we need to compare the maximum static friction force (fs\_{max}) with the gravitational force (mg) for the riders in the given mass range.}'' [Segment 2]  \\

Seeking coherence between ideas & ``{\em I need to figure out how much of their force is heading into the wall using their momentum momentum of the lightest person, but I don't know how to do that.}'' [Episode 3]  & ``[..] {\em we can see that fs\_max is significantly less than mg. This means that the maximum static friction force provided by the wall is not enough to counteract the gravitational force for a rider}[..]'' [Segment 3]   \\
\hline
\end{tabular}

\end{center}
\end{table*}

\subsubsection{Segment 1: Solutions to prompts A and B }

{\bf Summary.} The AI's response to the first prompt entails a total of five assumptions to solve the problem. The assumptions span across the forces acting on the rider, the rider's orientation on the wall, and the radius of the Gravitron. In response to the second prompt, the ChatGPT's response entails a free-body diagram as shown in Figure~\ref{fig:ai-fbd} along with it's description of gravitational, frictional, and centripetal forces.

\begin{quote}
{\em A. To solve this problem, we can make the following assumptions:

1. The gravitational force acting on the riders is the only significant force involved in determining their ability to stay against the wall. Other forces such as air resistance, normal force, or friction between the riders and the floor can be neglected.

2. The riders lean against the wall and do not apply any additional force to maintain contact with it. The only force keeping them against the wall is the static friction between their bodies and the wall.

3. The riders are evenly distributed along the wall of the cylinder, and their weight is uniformly distributed. This assumption allows us to consider the forces acting on a single rider instead of analyzing each individual rider separately.

4. The radius of the cylinder remains constant throughout the ride and does not change due to deformation or stretching.

5. The riders are motionless relative to the rotating frame of reference. In reality, there might be some slight movement or sliding, but for simplicity, we assume that they remain stationary with respect to the cylinder's rotating frame.

These assumptions simplify the problem and allow us to focus on the gravitational force and the static friction acting on a single rider.}

{\em B. When the room is spinning, we can create a free body diagram for the rider. In this case, the rider experiences two main forces: the gravitational force (mg) and the static friction force (fs) provided by the wall.
Here is the free body diagram:

In the diagram, the gravitational force (mg) acts vertically downward since it is the force due to the rider's weight. The static friction force (fs) acts radially inward, directed towards the center of the circular path. This force is responsible for providing the centripetal force required to keep the rider against the wall.

It's important to note that the static friction force adjusts its magnitude to exactly balance the gravitational force to prevent the rider from sliding down the wall. The maximum static friction force ($fs_max$) can be calculated using the formula $fs_{max} = \mu s \* N$, where $\mu s$ is the coefficient of static friction and N is the normal force (equal to the gravitational force in this case).

This free body diagram helps us understand the forces acting on the rider and how they contribute to keeping the rider against the wall while the room is spinning.}
\end{quote}

{\bf Features of sensemaking.} From the viewpoint of sensemaking, this segment reflects the blending of everyday and formal knowledge along with generation of ideas by describing the Gravitron mechanism in terms of formal concepts of physics. All five assumptions (in response to prompt A) and the description of the free-body diagram focusing on the interaction of various forces keeping the rider stay put evidence this feature. Furthermore, the segment also entails several conceptual and mathematical ideas of the involved forces.

{\bf Codes of mechanistic reasoning.} From the perspective of mechanistic reasoning, this segment entails the first six analytical codes. While the {\em description of the target phenomenon} is evidenced at the end of the description of the force diagram (``{\em [...] keeping the rider against the wall while the room is spinning.''}), the {\em set-up conditions} are identified in the beginning of the response to prompt B by identifying the forces acting on the rider when the room starts spinning. Several {\em entities} such as the forces, riders, radius of the cylinder, and coefficient of static friction are identified that play key roles in the observed phenomenon. In addition, the {\em activities} and {\em properties of entities} are highlighted through description of the forces acting on the rider (e.g., gravity activity downwards by virtue of the rider's weight). Lastly, this segment also captures the {\em spatial organization of various entities} such as the riders leaning and evenly distributed against the wall, constant radius of the Gravitron, static friction being directed towards the Gravitron's center, etc.

\begin{table*}
\begin{center}
\label{tab:mechanistic-reasoning-qualitative}
\caption{Representative evidentiary quotes highlighting the analytical codes of mechanistic reasoning in Joshua's and ChatGPT's responses to the scaffolded Gravitron task.}

\begin{tabular}{p{0.15\linewidth}p{0.36\linewidth}p{0.36\linewidth}}
\hline
Mechanistic codes & Joshua & ChatGPT \\
\hline 

Describing target phenomenon. & -  &  ``{\em To determine whether anyone will slide off the wall when the floor drops out} [..]'' [Segment 2]  \\

Identifying set up conditions. & -  & ``[...] {\em when the floor drops out at an angular velocity of 3 rad/s.}'' [Segment 3]  \\

Identifying entities. & ``{\em I have the centre of gravity. [..] It has to be angular momentum right?}'' [Episode 1]  & ``{\em The gravitational force acting on the riders is the only} [...]'' [Segment 1]  \\

Identifying activities.  & ``[..] {\em the force of friction is keeping them off the ground from Fg.}'' [Episode 2 ] & ``{\em The static friction force (fs) acts radially inward.} [...]'' [Segment 1]   \\

Identifying properties of entities. & ``[..] {\em I only have something pointing to the center of the circle but I don't know what's going out of the circle.}''  [Episode 2]  & `` [..] {\em the gravitational force (mg) acts vertically downward since it is the force due to the rider's weight.}'' [Segment 1]   \\

Identifying organization of entities. & ``{\em The angle of the wall... like.. when spinning should be orthogonal?}'' [Episode 1]  & ``{\em The riders are evenly distributed along the wall of the cylinder,} [...]'' [Segment 1]  \\

Chaining: backward and forward. & -  & ``[..] {\em  a rider weighing 160 kg would likely slide off the wall when the floor drops out.}'' [Segment 3]  \\

\hline
\end{tabular}
\end{center}
\end{table*}

\subsubsection{Segment 2. Solution to prompt C}

{\bf Summary.} In order to determine whether anyone will slide off the Gravitron's walls, the AI solution mentions of comparing the frictional and gravitational forces. Expressing friction in terms of normal force and equating (though incorrectly) the normal force to gravity, the solution argues that if the friction is greater than the gravity, then the riders would stay put. 

{\bf Transcript}
\begin{quote}
{\em C. To determine whether anyone will slide off the wall when the floor drops out at an angular velocity $\omega$ of 3 rad/s, we need to compare the maximum static friction force (fs\_{max}) with the gravitational force (mg) for the riders in the given mass range.

The maximum static friction force can be calculated using the formula fs\_max = $\mu$s * N, where $\mu$s is the coefficient of static friction and N is the normal force. In this case, the normal force is equal to the gravitational force acting on the rider, so N = mg.

If fs\_max is greater than or equal to mg, the riders will stay against the wall and not slide off. However, if fs\_max is less than mg, the riders will not have enough static friction force to prevent sliding and may potentially slide off the wall.}

\end{quote}

{\bf Features of sensemaking.} This segment of the AI transcript highlights the generation of ideas component of the sensemaking process. The arguments of relating the frictional and gravitational forces, and equating normal force to gravity evidence this feature.  

{\bf Codes of mechanistic reasoning.} This segment also highlights the {\em description of the target phenomenon, identifying set-up conditions}, and {\em chaining} codes of mechanistic reasoning. While the transcript's first sentence highlights the initial two codes (describing the target phenomenon and identifying set-up conditions), the final arguments, i.e., the necessary condition between the frictional and gravitational forces evidences the final code ({\em chaining}).

\subsubsection{Segment 3. Conclusion to prompt C}

{\bf Summary.} The condition is then mathematically formulated by equating the two forces and further simplifying the equation through substitution of the given values. In the end, the solution concludes that riders would slide off the Gravitron's walls, as the gravity outweighs the frictional force. 

\begin{quote}
{\em Let's calculate fs\_max for the riders at the upper end of the mass range, which is 160 kg. We'll assume the coefficient of static friction ($\mu$s) is 0.2, as stated in the initial question.

fs\_max = $\mu$s * N = $\mu$s * mg = 0.2 * 160 * 9.8 = 313.6 N

Now, we need to determine the magnitude of the gravitational force acting on the rider:
mg = 160 kg * 9.8 m/s\^2 = 1568 N

Comparing fs\_max (313.6 N) with mg (1568 N), we can see that fs\_max is significantly less than mg. This means that the maximum static friction force provided by the wall is not enough to counteract the gravitational force for a rider of 160 kg when the floor drops out at an angular velocity of 3 rad/s.

Therefore, with the given parameters, a rider weighing 160 kg would likely slide off the wall when the floor drops out. It's important to adjust the ride or implement additional safety measures to ensure riders' safety in the specified mass range.}
\end{quote}

{\bf Features of sensemaking.} This final segment evidences the coherence seeking feature of the sensemaking process. Making the final conclusion based on the calculated magnitudes of frictional and gravitational forces evidences this feature.

\begin{table*}
\begin{center}
\label{tab:sensemaking-quantitative}
\caption{Evidence of sensemaking elements in student and AI responses to the open-ended and scaffolded versions of the Gravitron task.}
\begin{tabular}{p{0.45\linewidth}p{0.1\linewidth}p{0.1\linewidth}p{0.1\linewidth}p{0.1\linewidth} }
\hline
\multirow{2}{*}{Sensemaking elements} & \multicolumn{2}{l}{Open-ended (8)} & \multicolumn{2}{l}{Scaffolded (8)} \\

  & Student & AI & Student & AI \\
\hline 

Noticing inconsistencies in understanding & 4  &  - & 4  & - \\

Use of everyday and formal knowledge & 6  & 8  & 8  & 8 \\

Generating and connecting ideas & 5  & 8  & 6  & 8\\

Seeking coherence between ideas & 4  & 8  & 6  & 8\\
\hline
\end{tabular}

\end{center}
\end{table*}

{\bf Codes of mechanistic reasoning.} On the other hand, this segment evidences {\em identifying the set-up conditions, the properties of entities, their activities,} and {\em chaining} codes of mechanistic reasoning. The initial set-up conditions are re-iterated through the phrase ``[...]{\em when the floor drops out at an angular velocity of 3 rad/s}. Furthermore, the numerical comparison between the magnitudes of the two forces evidences the code corresponding to identifying the properties of entities. Lastly, the concluding statement about the riders falling of the walls marks the identifying the activities of entities (i.e., riders) and {\em chaining} codes.  

\subsection{Comparative reflection between Joshua's and AI's solutions}

We examined a representative solution from a student (Joshua) and ChatGPT to the scaffolded  Gravitron task. The solution approaches were analyzed through the theoretical lenses of sensemaking elements and analytical codes of mechanistic reasoning (highlighted in Tables~\ref{tab:sensemaking-qualitative}~and~\ref{tab:mechanistic-reasoning-qualitative}). Both these approaches have distinct and distinguishing features from each other. 

Joshua's response reflects a student quitting a problem upon trying to solve it from multiple ways. Despite the unsuccessful attempt, the student's approach highlighted several key features of scientific inquiry including the meta-cognitive perception of discrepancies in one's own understanding, iterative refinement of arguments, and leveraging representation-based reasoning. Joshua perceived an inconsistency in his understanding about the forces acting on the Gravitron's riders. He sought to address this gap by through identifying several forces and trying to recollect his knowledge about centripetal forces. In the process, he iteratively refined his arguments, initially considering the construct of angular momentum, then exploring force acting against centripetal force (centrifugal force), and later reflecting on impulse and the mathematical definition of a force. These shifts in arguments were accompanied by representations such as the constructed free-body diagram in Figure~\ref{fig:joshua-solution} and gestures as shown in Figure~\ref{fig:joshua-gesture}. 

On the other hand, the ChatGPT's solution completed the problem through requisite prediction. Despite accompanying conceptual  flaws (e.g., equating normal force to gravity) and providing incorrect conclusion, the solution reflected an elegant structure clearly highlighting the underlying assumptions necessary to solve the problem, the target rider-Gravitron system, relevant forces acting on the rider, and a descriptive conclusion. This clear articulation of various facets of the solution resulted in evidencing all the codes of mechanistic reasoning thereby reflecting mechanistic sophistication.     

\section{Quantitative distribution of sensemaking and mechanistic reasoning features across the two data sets}
\label{sec:quantitative}

In this section, we detail the features of sensemaking and mechanistic reasoning manifesting in students and AI responses across the two versions of the Gravitron task. We first discuss the sensemaking elements before moving on to the analytical codes of mechanistic reasoning. 

\subsection{Sensemaking}

Table~\ref{tab:sensemaking-quantitative} highlights the sensemaking elements evidenced in students' and AI's solutions across the two versions of the Gravitron task. 

Unlike student solutions, all of the AI responses exhibited the sensemaking features with the exception of noticing ``knowledge gaps''. Around 50\% of the student responses evidenced the metacognitive feature of noticing inconsistencies in their knowledge systems (similar to the one's observed in Joshua's case discussed in the previous section). However, though not surprising, none of the AI responses evidenced this feature. At least as of writing this manuscript, large language models have not been observed to notice gaps in ``their knowledge systems'', a crucial feature of the sensemaking process. 

Concerning the second sensemaking element, blending everyday and formal knowledge, six of the eight student solutions to the open-ended task and all eight responses to the scaffolded version evidenced conceptual reasoning about the given system. The two solutions in the open-ended version did not reflect this feature, primarily due to the fixation on ``choosing the right equation'' and ``plugging-and-chugging'' of equations. However, a students' better evidencing of this sensemaking feature to the scaffolded version may be due to the presence of sub-prompts, where the participants were first asked to identify the forces (and assumptions) acting on the rider before solving the problem. Explicit identification of the forces made students to describe the features of the Gravitron's ride in terms of the formal concepts of physics thereby evidencing the second sensemaking element. On the AI front, all responses, irrespective of the problem format were explicit in describing the interplay of various concepts in the functioning of Gravitron system. This feature can also be seen in the sample solution analyzed in the previous section.  

On the third sensemaking element -- generating and connecting ideas -- the sophistication of AI responses across the two problem formats seemed to be relatively better as compared to student ones. This could be largely attributed to AI's varied assumptions and detailed descriptions of mathematical formulae in each solution. In contrast,  the student responses tended to have less detailed formulaic explanations, which might be a result of having access to an equation sheet during the interview. Secondly, unlike AI responses, students' arguments reflected iterative refinement of ideas - a key feature of productive learning in science. Though it's not surprising that AI responses did not iteratively refine arguments, there were instances in solutions where certain conceptual arguments made in the earlier phases of solution were not followed up.   

On the penultimate feature of sensemaking, i.e., seeking coherence between the generated ideas, all AI solutions reflected this feature whereas, about half and three-fourth of the student solutions each across the two formats reflected a coherent conclusion. The last sensemaking element, generating mechanistic accounts is detailed below.

\begin{table*}
\begin{center}

\label{tab:mechanistic-reasoning-quantitative}
\caption{Distribution of mechanistic reasoning codes in student and ChatGPT responses to the open-ended and scaffolded versions of the Gravitron task.} 

\begin{tabular}{p{0.47\linewidth}p{0.09\linewidth}p{0.09\linewidth}p{0.09\linewidth}p{0.09\linewidth} }
\hline
\multirow{2}{*}{Markers of mechanistic reasoning} & \multicolumn{2}{l}{Open-ended (8)} & \multicolumn{2}{l}{Scaffolded (8)} \\

  & Student & AI & Student & AI \\
\hline 

Describing target phenomenon. & 8  &  8 & 8  & 8 \\

Identifying set up conditions. & 8  &  8 & 8  & 8 \\

Identifying entities. & 8  &  8 & 8  & 8 \\

Identifying activities. & 7  &  8 & 8  & 8 \\

Identifying properties of entities. & 7 &  8 & 8  & 8 \\

Identifying organization of entities. & 6  &  8 & 8  & 8 \\

Chaining: backward and forward. & 5  &  8 & 6  & 8 \\

Use of diagrams & 8  & 3  & 8  & 8 \\

Use of equations & 8  &  8 & 8  & 8 \\

Use of gestures & 4  & -  & 4  & -\\
\hline
\end{tabular}

\end{center}
\end{table*}

\subsection{Mechanistic Reasoning}
   
Table~\ref{tab:mechanistic-reasoning-quantitative} highlights the analytical codes of mechanistic reasoning evidenced in student- and AI-generated solutions across the open-ended and scaffolded formats of the Gravitron task. Overall, AI responses evidenced sophistication in mechanistic reasoning (albeit providing incorrect arguments) whereas student solutions reflected the effective use of representations (diagrams and gestures).

One of the crucial distinctions between the two sets of responses (also evident from the two sample solutions analyzed in the previous section) lie in evidencing the lower and higher codes, particularly {\em describing the target phenomenon, identifying set up conditions, and backward/forward chaining}. Many students did not explicitly describe the lower codes such as the target phenomenon or the initial set up conditions of the Gravitron. These were rather inferred from their interaction with the force diagrams. The absence of lower codes may be perhaps due to students considering such arguments to be ``obvious'', especially considering they had already solved around five problems prior to the Gravitron task. Furthermore, eleven student responses across the two problem formats did not evidence the higher hierarchical code of {\em chaining}, primarily because of not being able to solve the problem completely and make the required prediction.  Additionally, Table~\ref{tab:mechanistic-reasoning-quantitative} also highlights explicit scaffolding to be influential in evidencing the higher codes.     

Next crucial difference between the two sets of responses corresponded to use of representations in mechanistic reasoning reasoning. While every student solution to the open-ended problem accompanied force diagrams, only three of the AI solutions reflected the same. Though a contemporary shortcoming, AI generated diagrams were relatively unsophisticated (as can be seen in Figure~\ref{fig:ai-fbd}) and were not referred in the solutions.  

\section{Discussion and Implications}
\label{sec:discussion}

In the previous two sections, we examined student- and AI-generated solutions to a physics problem posed in two different formats. The solutions were analyzed through the theoretical lenses of sensemaking and mechanistic reasoning. While Section~\ref{sec:qualitative} presented qualitative analysis of representative solutions, Section~\ref{sec:quantitative} presented the quantitative distribution of the processes across each data set. The results, highlighted in Tables~\ref{tab:sensemaking-qualitative},~\ref{tab:mechanistic-reasoning-qualitative},~\ref{tab:sensemaking-quantitative}~and~\ref{tab:mechanistic-reasoning-quantitative}, can be summarized as:

\begin{enumerate}
    \item The AI solutions evidenced most features of the sensemaking process and all codes of mechanistic reasoning across both open-ended and scaffolded versions of the Gravitron task. The student responses on the other hand, evidenced a subset of the sensemaking features and most codes of the mechanistic reasoning.
    
    \item AI responses were particularly better at describing the physical features of the given system in light of conceptual ideas (blending everyday and formal knowledge) and evidencing the lower and higher codes of mechanistic reasoning.
    
    \item Half of the total student responses highlighted the feature of noticing gaps in their understanding and iterative refinement of arguments in their solutions. 

    \item Unlike AI responses, student solutions effectively leveraged diagrams in making sense of the given context. 
    
    \item While scaffolding had no effect on AI solutions, the student responses better evidenced the mechanistic reasoning codes when scaffolded with explicit prompts. 
\end{enumerate}

Across each data set, all AI solutions explicitly evidenced the seven codes of mechanistic reasoning. In contrast, student responses (e.g., Joshua's response in Section~\ref{sec:qualitative}), did not explicitly evidence the lower and higher analytical codes, particularly the ones pertaining to calling out the target phenomenon, describing the system's initial conditions, and predicting the system's behavior based on its current state. On the sensemaking front, the AI solutions also clearly described the rider-Gravitron system in terms of physics concepts (blending everyday and formal knowledge) such as the interplay of various forces on the riders and the conditions necessary for sustaining riders inside the Gravitron.

Another distinguishing feature between the two sets of responses corresponded to noticing gaps in knowledge system. Half of the total students evidenced noticing inconsistencies in their understanding surrounding the knowledge required to solve the Gravitron task. This meta-cognitive reflection resulted in students iteratively refining their arguments as they progressed in their approaches. On the other hand, none of the AI responses reflected this feature. This should not come as a surprise as the large language models are not capable (at least as of writing this manuscript) of noticing gaps in their ``knowledge system''. This result however needs to be reemphasized as the interactive, grammatically sound, assertive, and expert-like answers can make one to associate the probabilistic responses generated by an algorithm to a human ``responding from the other end''. As Kortemeyer~\cite{kortemeyer2023could} observes

\begin{quote}
``{\em It is irritatingly hard not to anthropomorphize ChatGPT. As a physics teacher, one invariably finds oneself rooting for the students and thus by extension also for ChatGPT, celebrating its successes and being frustrated about its occasionally inexplicable failures.}''    
\end{quote} 

Another unsurprising but distinguishing feature between the two sets of responses corresponds to the ways of leveraging representations in the problem-solving exercise. Every student who engaged with the problem constructed a diagram and referred to it several times during their reasoning. However, only 3 AI solutions generated representations without being specifically prompted to do so in the open-ended task. Even the ones which accompanied free-body diagrams (Figure~\ref{fig:ai-fbd}) were relatively unsophisticated and were not consistently referred in the solutions.  This is again not surprising as generation of sophisticated representations is a major shortcoming of contemporary large language models.    

We also observed explicit prompting in problems, particularly those involving diagram-based reasoning, to elicit more sophisticated mechanistic accounts from students. Most of the participants evidenced the higher codes of mechanistic reasoning while responding to the scaffolded version. Perhaps, this could be attributed to the requirement in the scaffolded task for participants to first reason about the interaction of forces on the rider before making their prediction. In the context our problem, explicit emphasis on forces results in evidencing several codes related to entities (properties, activities and organization of entities). However, there was no dependence observed between the codes and the problem format in ChatGPT's responses.   

Lastly, though beyond the purview of the two processes, AI solutions often included a variety of assumptions necessary for their conclusions to be valid. In the open-ended version, without specific prompting, three out of eight solutions explicitly stated the assumptions underlying their conclusions. On the student front, students hardly referred to assumptions when solving open-ended task, and when explicitly prompted, they often struggled to articulate assumptions~\cite{sirnoorkarassumptions2023}. Secondly, none of the AI responses across the two problems provided a correct solution despite exhibiting sophistication on various aspects. This observation aligns with findings from other studies which have made cautioned about ChatGPT's seemingly confident solutions despite accompanying major conceptual flaws~\cite{kortemeyer2023can,yeadon2023exploring}. 

The above-mentioned results have several implications for educators in leveraging AI for teaching and learning of physics. Firstly, our results highlight that AI is particularly effective at conveying the vocabulary of physics (mechanistic codes such as highlighting the interplay between physical features, mathematical formalisms, and conceptual ideas). This vocabulary is further embedded in a systematically organized solution structure such as initial descriptions of target systems, mathematical calculations, and clear articulation of conclusions accompanied by underlying assumptions. These aspects are often  beneficial for beginners in early stages of their physics courses in effectively structuring and communicating their results. Activities such as and multilingual students in the early phases of their courses. asking students to solve challenging problems or write descriptive essays (say about, fundamental laws of physics), and making them compare their responses to AI-generated ones can tactfully communicate the physics terminology along with the structural communication of ideas. Such practices can also be useful for multilingual students getting started with learning physics in English language.

Furthermore, there are ongoing efforts in designing large language models (LLMs) that can provide personalized learning assistance to students~\cite{extance2023chatgpt}. Our observation of explicit prompting to facilitate students' mechanistic reasoning can lead to designing planned scaffolds focused on the analytical codes that can support students in generating sophisticated mechanistic accounts. Calibrated interventions aimed at nudging students towards identifying the appropriate target system, relevant entities, and associated activities can lead to enhanced learning about real-world systems, particularly in university-level physics. For instance, in our case study, an explicit scaffold asking Joshua to reconsider the relevance of inward directed force or the role of normal force in providing the necessary centripetal force would have helped him in continuing the problem-solving endeavor.  In a similar way, several studies are also exploring on providing personalized feedback based on students' solutions. The features of our adopted frameworks (sensemaking elements and analytical codes of mechanistic reasoning) can also inform the nature and structure of feedback.    

Consistent with the contemporary literature, we too observed AI responses confidently articulating incorrect solutions through sophisticated vocabulary. This feature can be leveraged by asking students to critique the AI-generated solutions. Such exercises can not only make students aware of the imperfections underlying AI responses, but can also implicitly communicate the sophisticated lexical patterns of scientific communication. In addition, confident phrasing of incorrect solutions can also be a fertile ground for students to notice gaps in their knowledge systems, thereby facilitating sensemaking in our classrooms. 

Contemporary studies have also called for designing next-generation assessments in which Generative-AI can support students instead of impeding their learning. Our results highlight designing assessments focused on diagram- and gesture-based reasoning to be potentially productive as the contemporary AI systems seem to be relatively less unsophisticated in responding to such prompts.

\section{Conclusion, Limitations, and Future Work}
\label{sec:conclusion}

The recent advent of Generative-AI, particularly ChatGPT, has garnered significant traction in physics education community. A predominant section of studies have focused on gauging the accuracy of AI in answering various forms of physics assessments. We contribute to these studies by going beyond capturing the correctness of AI-generated responses and comparatively examining them with students' reasoning to two versions of a physics problem based on real-world context. Using mixed-methods approach, we analyze the responses through the lens sensemaking and mechanistic reasoning - the processes central to scientific inquiry. We observed that while AI responses evidenced most features of the two processes through well-structured solutions, student responses effectively leveraged representations in their solutions through iterative refinement of arguments.  In other words, while AI responses reflected the sophisticated lexical patterns of experts' vocabulary (albeit providing incorrect solutions), the student approaches reflected the rich epistemic practices of physics. That is, while the AI responses reflected how {\em physics is talked about}, the student responses reflected {\em how physics is practiced}. 

Engaging with physics not only entails leveraging diverse arguments such as experimental facts,  approximations, etc.~\cite{sirnoorkar2020towards,sirnoorkar2016students}, but also entails communicating ideas and results using appropriate vocabulary. This vocabulary may entail clear description of the target system under investigation, discussion of the relevant physical quantities and mathematical formalisms leveraged in the given context, or providing appropriate justification for conclusions (particularly in the context of problem solving)~\cite{bergeron2021towards,ACER}. ChatGPT consistently mirrored these features through well-structured responses, thereby reflecting {\em how physics is talked about}.   

On the other hand, student responses effectively reflected the conventional knowledge-building practices in physics, or {\em how physics is practiced}. These included representation-based reasoning (use of diagrams and gestures), along with iterative refinement of arguments. Diagram- and gesture-based reasoning are central to knowledge creation in physics. Conceptual ideas such as the direction of magnetic field around a current carrying conductor or the distribution of electric-field lines around a system of charges are often discussed through gestures and diagrams. Furthermore, conventional knowledge-building practices in physics such as argumentation and modeling are iterative in nature, i.e., these practices involve iterative refinement of arguments in making sense of novel contexts~\cite{bing2012epistemic,passmore2014models}.  

This study makes several contributions to the contemporary literature focused on the intersection of artificial intelligence in STEM education. Firstly, this study goes beyond gauging the accuracy of AI responses to analyzing them through the lens of cognitive processes that are central to scientific inquiry. We adopt a framework-based approach in unpacking the subtle features of scientific inquiry in AI generated solutions and comparing them with students' responses. Secondly, though several studies have explored ChatGPT's causal reasoning in diverse domains such as neuropathic pain diagnosis~\cite{tu2023causal} and publicly available multilingual multi-modal data~\cite{bang2023multitask}, this study explores mechanistic reasoning (which subsumes causal reasoning) in physics with a focus on diagrams and equations.     

Our findings however accompany several limitations. Firstly, the claims made in this study are based on the analysis of limited number of student and AI responses to just one physics problem. Analysis of a larger data set of responses to a diverse physics problems would undoubtedly shed more insights. Secondly, the students interview involved interventions from the interviewer during prolonged pauses and may have potentially impacted the participants' ``chain of thought''. We acknowledge this inherent limitation of think-aloud interview and in our approach to data collection. Finally, the data collected from ChatGPT and that from students are categorically of different nature. While students' data consisted of verbal arguments, gestures, and written solutions, ChatGPT's data consisted entirely of texts. Comparing textual data across the two would have been an ideal scenario. However, our primary objective was to explore how students naturally reason about a context and how AI would respond to the same context. As practitioners, we were aware that students would not prefer to write every argument in their solutions, particularly to as complex problem as the one described in this paper. Thus, we acknowledge that a major limitation of this work corresponds to making claims by analyzing two categorically different forms of data.

Future work would involve analyzing AI and student solutions to other problems from our data through the lens of additional processes such as modeling and argumentation that are integral to scientific reasoning. Further studies would also focus on extending similar methodology across students' written essays and AI responses. Analysis of AI- and student-generated assumptions along with the epistemological messages conveyed by AI responses to students on what counts as ``knowing'' or ``doing'' physics are also avenues for future explorations. 

\section{Acknowledgements}
Special thanks to Katherine C. Ventura and Amali Priyanka Jambuge for creating the assessments and conducting the interviews reported in this study. 

\section{Declaration of generative AI in scientific writing
}
During the preparation of this work the first author used ChatGPT-4 in order to improve language and readability. After using this tool/service, the author reviewed and edited the content as needed and takes full responsibility for the content of the publication.
 \bibliographystyle{elsarticle-num} 
 \bibliography{references}





\end{document}